\documentclass[11pt]{article} 
\usepackage{amssymb}
\usepackage{graphicx,epsf,rotate,subfigure} 
\usepackage{times,cite,color}
\usepackage{multirow}
\usepackage{amsmath}
\begin{document}

\begin{flushright}
RECAPP-HRI-2013-006
\end{flushright}

\vskip 30pt

\begin{center}
{\Large \bf Constraining minimal and non-minimal UED models with Higgs
  couplings} \\
\vspace*{1cm} \renewcommand{\thefootnote}{\fnsymbol{footnote}} { {\sf
    Ujjal Kumar Dey${}^{1,2}$}\footnote{email: ujjaldey@hri.res.in}
  and {\sf Tirtha Sankar Ray${}^{3}$}}\footnote{email:
  tirtha.sankar@unimelb.edu.au} \\
\vspace{10pt} {\small ${}^{1)}$ {\em  Harish-Chandra Research Institute,
    Chhatnag Road, Jhunsi , Allahabad 211019, India} \\ ${}^{2)}$ {\em 
    Department of Physics, University
    of Calcutta, 92 Acharya Prafulla Chandra Road, Kolkata 700009,
    India}\\ ${}^{3)}$ {\em
    ARC Centre of Excellence for Particle Physics at the Terascale,
    School of Physics, \\University of Melbourne, Victoria 3010,
    Australia}} \normalsize
\end{center}

\begin{abstract}
Early indications from the LHC for the observed scalar boson imply
properties close to the Standard Model Higgs, putting considerable
constraints on TeV scale new physics scenarios. In this letter we
consider flat extra dimensional scenarios with the fifth spatial
dimension compactified on an $S^1/Z_2$ orbifold.  We find in the
minimal model the experimentally preferred effective Higgs couplings 
to gluon and photon at $95\%$ confidence level disfavor the New Physics 
scale below $1.3$ TeV. We demonstrate that a generalization of these models to include
brane localized kinetic terms can relieve the tension to accommodate
scales as low as $0.4$ TeV.

\vskip 5pt \noindent
\texttt{Key Words:~Extra dimension}
\end{abstract}

\textbf{Introduction:} Increasing evidence shows that the observed
scalar boson at CERN as reported by ATLAS \cite{:2012gk} and CMS
\cite{:2012gu} collaborations has properties that are close to the
Standard Model (SM) Higgs.  The era of precision measurements at LHC and
future colliders that lie ahead will sit on judgment on the Higgs
properties.  However present trends indicate considerable constraints
on New Physics (NP) scenarios that imply a modification of the Higgs
production cross sections and branching ratios.  Of particular
importance are the loop driven Higgs couplings to the photon and the
gluon that are susceptible to considerable corrections from TeV scale
new physics.  Indeed the impact of these couplings has been explored
extensively in the recent past \cite{Buckley:2012em}.  Aficionados of
TeV scale new physics scenarios have been compelled to move to more
general versions of specific models typically with a larger parameter
space to accommodate these experimental constraints.
 
 In this letter we focus on compactified extra dimensional models with
 a flat bulk profile \cite{Anto:1990}.
 These models have enjoyed
 considerable attention owing to their minimal nature, prediction of
 natural dark matter candidate \cite{Servant:2002aq} and ability to
 mimic supersymmetric extension of the SM at the collider
 \cite{Datta:2005zs}, among other features.  In their simplest avatar,
 the Universal Extra Dimension (UED) models extend the SM with two new
 parameters viz. (i) the size of the compactified extra dimension
 $(R)$ and (ii) the cut-off of the effective theory $(\Lambda)$.
 The effect of predicted new states, the so called Kaluza-Klein (KK)
 excitations of the SM fields, on the Higgs couplings have been
 studied in \cite{Petriello:2002uu}. Corresponding studies in the
 context of warped 5d scenarios with explicit KK parity
 \cite{Bhattacharyya:2009nb} or KK parity violating custodial
 models\footnote{Note that in custodial Warped models, the sign of the
   contribution depends on the order in which the Higgs is regularized
   to the IR brane and the KK sum is made, see
   \cite{Carena:2012fk}. }\cite{Azatov:2010pf} have also been made.

In this letter we discuss the constraints from the LHC Higgs results
on the UED scenario, which we find to be quite severe. However
generalizing to models with brane localized kinetic terms (BLKT)
\cite{Flacke:2008ne} that arise naturally due to the cut-off
dependent radiative corrections, facilitates a considerable recovery
of the constrained parameter space.

This letter is organized as follows. We begin by reviewing the
formalism to study the Higgs couplings $gg\rightarrow h$ and
$h\rightarrow \gamma \gamma$, including possible contributions from
new states beyond the SM. Then we compute in turn the contributions
from UED and BLKT scenarios and compare with present experimental
values.  Finally we conclude.

\textbf{The loop driven Higgs couplings:} At the LHC, the Higgs
production chiefly proceeds through the gluon fusion process $gg
\rightarrow h,$ driven by the fermion triangle loop
\cite{Djouadi:2005gi}.  Within the SM the top loop contribution
dominates. Similarly an important decay mode for a 125 GeV Higgs is
the di-photon channel. This proceeds through a fermion and a W boson
loop within SM. New states with correct quantum numbers can show up as
virtual particles in these loops and may lead to a sizable correction
of the effective couplings.  The corresponding cross section and decay
width, including possible contributions from new massive states are
given below \cite{Cacciapaglia:2009ky},
\begin{eqnarray}
\Gamma_{h \rightarrow \gamma \gamma} &=& \frac{G_F \alpha^2 m_H^3}{128
  \sqrt{2} \pi^3} \left| A_W (\tau_W) + 3 \left( \frac{2}{3} \right)^2
A_t (\tau_t)\; + N_{c,NP} Q_{NP}^2\ {\mathcal{A}}_{NP}(\tau_{NP}) \right|^2\,,
\nonumber \\ \sigma_{gg \rightarrow h} &=& \frac{G_F \alpha_s^2 m_H^3}{16
  \sqrt{2} \pi^3} \left| \frac{1}{2} A_t (\tau_t)\; + C (r_{NP})
{\mathcal{A}}_{NP}(\tau_{NP})\ \right|^2\,,
\label{mastereqn}
\end{eqnarray}
where,
\begin{equation}
 {\mathcal{A}}_{NP}(\tau_{NP}) = 
\sum_{NP}\;  
  \frac{v}{m_{NP}} \frac{\partial m_{NP}}{\partial v}\;
A_{F,V,S} (\tau_{NP}). 
\label{np}
\end{equation}
Note that the contribution of a state in the loops is dependent on
the spin statistics of the new states ($F,V$ and $S$ stand for fermions,
vector bosons and scalars respectively). The various contributions are
given as,
\begin{eqnarray}
 A_F (\tau) &=& \frac{2}{\tau^2} \left( \tau + (\tau-1) f (\tau)
 \right)\,,~~ A_V (\tau) = - \frac{1}{\tau^2} \left( 2 \tau^2 + 3 \tau
 + 3 (2 \tau - 1) f (\tau) \right),\,\nonumber \\ A_S (\tau) &=& -
 \frac{1}{\tau^2} \left( \tau - f (\tau) \right)\,,
\end{eqnarray}
where, $\tau_i = m_h^2/4m_i^2,$ and $f (\tau) = \arcsin^2\sqrt{\tau}.$
In the light Higgs limit i.e., $m_h << 2 m_i,$ we get, $ A_F  \sim
{4}/{3}, A_V \sim - 7, A_S \sim {1}/{3}.$

It will be useful to define two dimensionless parameters, $ C_{gg} =
\sigma_{gg \rightarrow h}^{NP}/\sigma_{gg \rightarrow h}^{SM}$ and $
C_{\gamma\gamma} = \Gamma_{h \rightarrow \gamma \gamma}^{NP} /
\Gamma_{h \rightarrow \gamma \gamma}^{SM},$ to compare the correction
induced by New Physics scenarios with experimentally allowed
values. The best fit values for $C_{gg}$ and $C_{\gamma \gamma}$ from
the LHC Higgs results were computed in \cite{Buckley:2012em}. We are
going to use the numerical values quoted in \cite{Ellis:2013lra}:
$\sqrt{C_{gg}} =0.88 \pm 0.11 $ and $\sqrt{C_{\gamma \gamma}} = 1.18 \pm 0.12.$

\textbf{Universal Extra Dimension:} At lowest order, flat extra
dimensional theories like the UED \cite{Anto:1990} with the extra
spatial dimension compactified over a $S^1/Z_2$ orbifold, is a simple
extension of the SM with a single new parameter, the radius of
compactification $(R)$ of the extra spatial dimension added to the
theory. As 5d quantum field theories are essentially
non-renormalizable, physical parameters are sensitive to the finite
cut-off $\Lambda$ of the theory.  The end points $(x_5 = \pm \pi R/2)$
of the compactified extra dimension are the locations of four
dimensional hyper-surfaces called the 3-branes. The SM can be embedded
in the 5d scenario, the relevant part of the 
action be schematically represented by the following,

\begin{eqnarray}
S &=& \int_{-\pi R/2}^{\pi R/2} dx_5 \int d^{4}x \, \Bigg\{ \,
-\frac{1}{2} \sum_{i=1}^{3} {\rm Tr} \left[ F_{i MN}F_{i}^{MN} \right]
+ (D_{M}H)^{\dagger} D^{M}H + \mu^2 |H|^2 -\frac{\lambda_5}{4} |H|^4
\Bigg.\nonumber \\ & & \Bigg. + i \, \bar{Q}_3 \!\not\!{D} Q_3 + i \,
\bar{u}_3 \!\not\!{D} \, u_3  - \left[ i
  \lambda^{u}_{5} \overline{Q}_3 u_3 \tilde{H}  +{\rm h.c.}  \right] \Bigg\} \,\, ,
\label{uedaction}
\end{eqnarray}
where, the Latin index runs from $0$ to $3$ and $5$
(representing the flat extra dimension) the covariant
derivative is defined as $D_{M} = \partial_{M} - i\sum_{i=1}^{3}
g^{i}_{5}\, T^{a}_{i} A^{a}_{i\,M}$ and $\Gamma^{M} = \{ \gamma^{\mu},
i\gamma^5\}.$  The leptons can be
included analogous to the quarks, which we omit for brevity.  The 4d
effective action can be derived by performing a Fourier expansion of
the fields as a function of the compactified fifth dimension and
integrating out the fifth dimension term by term. This procedure
elevates every five dimensional bulk field to an infinite tower of
increasingly massive states. Depending on the boundary values at the
locations of the orbifold fixed points, the lowest lying states can be
massless (zero modes) and are identified with the SM fields. The
respective KK excitations, corresponding to higher Fourier modes, form
an infinite tower of massive states. The tree level masses of the $n$-th
KK excitation are given by ,

\begin{equation}
  (m^{(n)})^2 = (m^{(0)})^2 + n^2/R^2.
  \label{massued}
 \end{equation} 
For all the SM fields the zero mode mass can be written in the form,
$m^{(0)} \propto gv$, where $g$ is the gauge coupling for massive
vector bosons, the Yukawa coupling for fermions and the self coupling
for the Higgs, and $v$ is the vacuum expectation value of the zero
mode Higgs. The integer $n$ corresponds to quantized momentum along
the fifth dimension. 5d Lorentz invariance implies that every vertex
preserves the quantum number $n.$ However, orbifolding required to
obtain chiral fermions at the zero mode breaks this symmetry, leaving
a residual reflection symmetry in the 5d geometry. An important
aspect of the symmetric nature of the fifth dimension is that the KK
excitations have identical couplings as their zero modes and at every
vertex KK parity is preserved.

 We are interested in the contribution of the KK excitations that may
 show up as virtual particles and modify the loop level coupling of
 the zero mode scalar field that is identified with SM Higgs
 field. The contribution from the entire KK top tower is a convergent
 sum and can be computed in closed form in the light Higgs
 approximation by using Eqs.~\ref{np} and \ref{massued} and is given by
 \cite{Cacciapaglia:2009ky},
 \begin{equation}
 {\mathcal{A}}_t^{NP} \sim \frac{8}{3}\left(\frac{\pi m_t/R \coth(\pi m_t/R)
   -1}{2} \right).
 \end{equation}
  In the case of the KK version of the W boson loop, one needs to be
  careful to also incorporate the contribution from the additional
  scalar states present in the spectrum. These are states that are a
  linear combinations of KK excitations of the fifth component of the
  gauge bosons $W_5^{(n)}$ and the KK excitations of the Goldstone
  modes $G^{(n)},$ that are orthogonal to the longitudinal component
  of the massive $W_{\mu}^{(n)}$ boson. A systematic study of this
  within the light Higgs approximation gives us,
 \begin{equation}
  {\mathcal{A}}_W^{NP} \sim - \frac{20}{3} \left(\frac{\pi m_W/R \coth(\pi m_W/R)
    -1}{2}\right).
  \label{uedw}
 \end{equation}

\begin{figure*}
 \centering
 \subfigure{\includegraphics[width=.49\textwidth,keepaspectratio,angle=0]
   {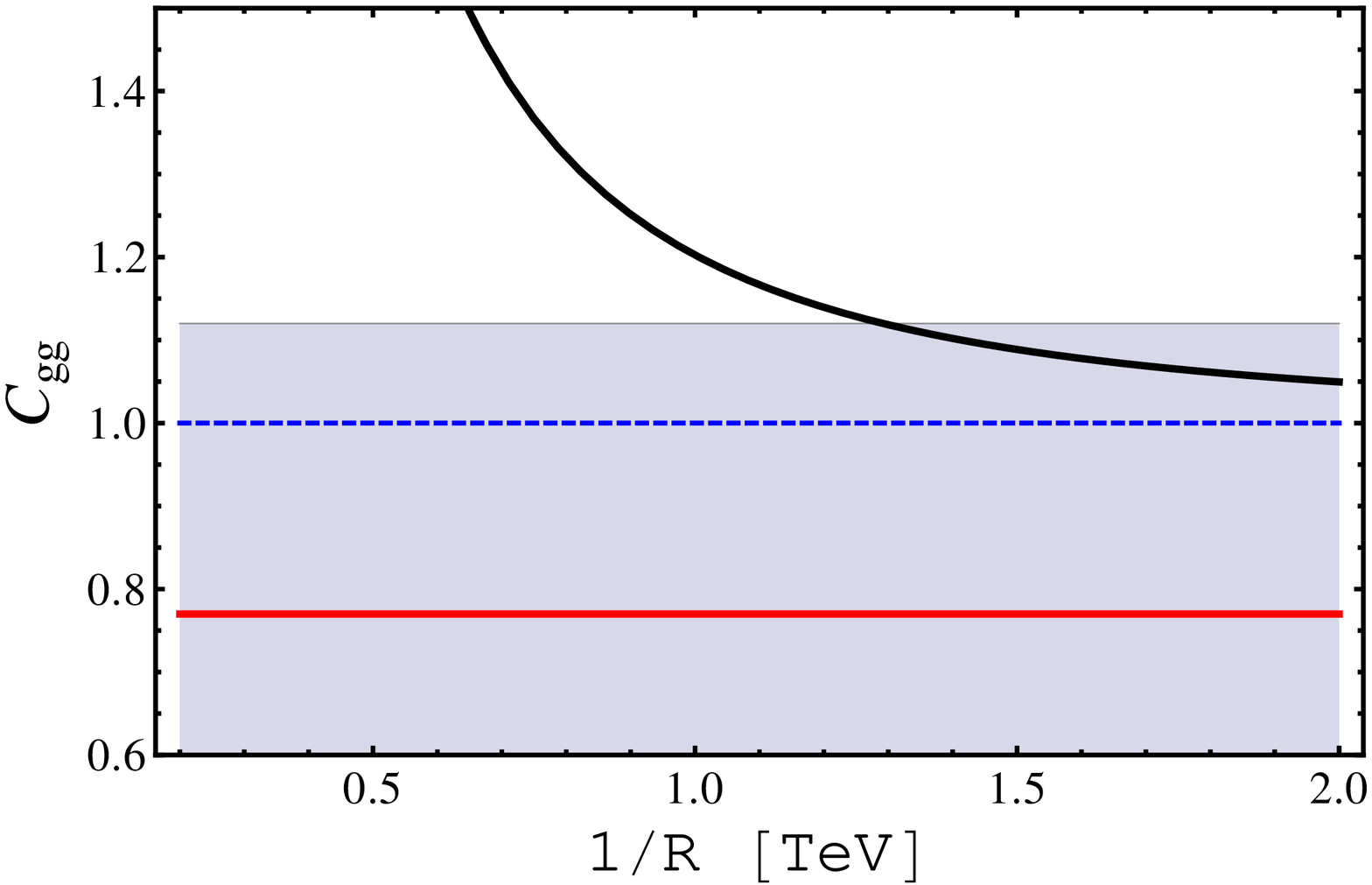}}\subfigure{\includegraphics[width=.49\textwidth,keepaspectratio,angle=0]
   {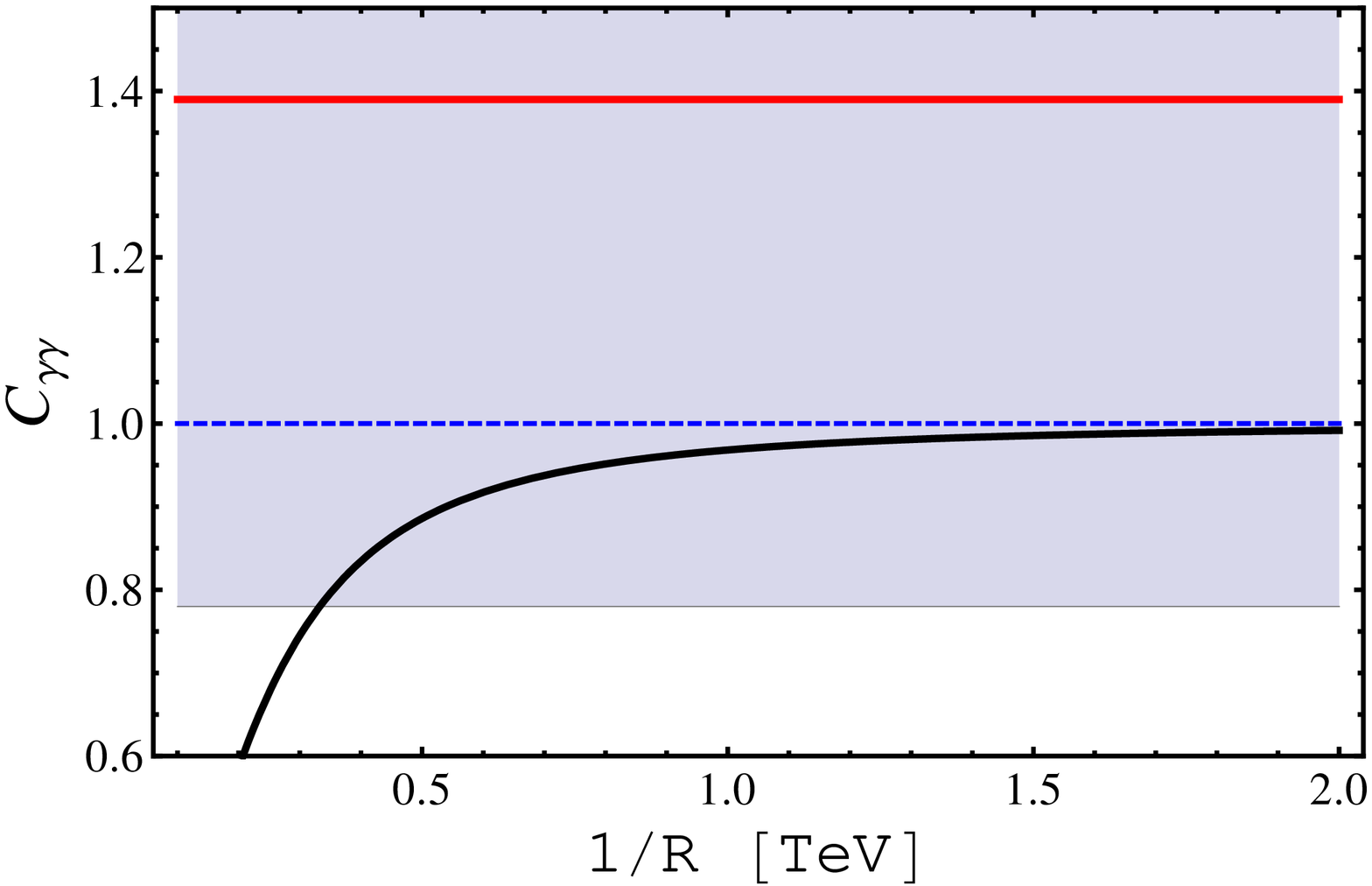}}
             \caption{ The ratios of the Higgs couplings in UED scenario
               to their SM values are plotted as  functions of the
               inverse radius of compactification of the extra
               dimension (1/R). The gray (shaded) bands represent the
               95\% confidence level best fitted values for these
               ratios, with the red (light) horizontal lines
               representing the central values from LHC data. The blue
               (dashed) lines correspond to the SM points. The black
               (dark) curves are the UED predictions. We have assumed $m_h = 125$ GeV.}
            \label{ued}
\end{figure*}

 Using these expressions in Eq.~\ref{mastereqn}, we compute the
 correction to the Higgs couplings as a function of the radius of
 compactifications of the fifth spatial dimension. Figure~\ref{ued}
 shows that the measurement of the 
 Higgs couplings to gluons  at 95\% confidence disfavors $1/R < 1.3$ TeV.
 The constraints from
 the effective couplings to photons are smaller primarily owing to the
 partial cancellation of the contributions between the KK fermions and
 KK gauge-Goldstone bosons. This should be compared with other
 constraints which are typically strongest from the oblique
 electroweak corrections estimated to be around: $1/R > 0.8$ TeV
 \cite{Belyaev:2012ai}. The bounds from direct searches at colliders
 typically lead to sub-dominant or comparable constraints
 \cite{Edelhauser:2013lia}, owing to the relatively compressed mass
 spectrum within this class of models. Recent LHC Higgs mass bounds can 
 constrain UED from the renormalization group running of the 
 physical parameters \cite{Blennow:2011tb}. For constraints from vacuum 
 stability bounds see \cite{Datta:2012db}.
 It may be noted
 that an exclusion limit at $1/R > 1.3$ TeV closes in on the
 overclosure limit from dark matter relic abundance in the minimal
 model \cite{Belanger:2010yx}. Admittedly this limit may be avoided
 by reorganizing the mass spectrum in non-minimal models, as
 demonstrated in \cite{Kong:2005hn}.
 
 Thus we find that if the current trend of the Higgs data gets support in
 future with increased statistics, it will result in tighter bounds
 than from direct observations and pose a challenge to the parameter
 space of these models that is accessible to present and future
 collider experiments.

\textbf{Brane Localized Kinetic Terms:} In higher dimensional theories
with lower dimensional defects, one expects localized divergent terms
at the location of the defects.  This arises due to the violation of
translation invariance at these locations leading to UV sensitive
radiative corrections that run with the scale and cannot be removed
simultaneously at all scales. A way to quantify these corrections is
to consider an effective theory at the cut-off scale, with higher
dimensional operators that are allowed by the standard model gauge
symmetries and Lorentz invariance to be added to the 5d theory. Such
operators can be in the bulk or localized at the orbifold fixed
points. Purely from the universality class that they belong to the
brane localized kinetic terms, being relevant operators due to their
mass dimension, lead to the most important corrections at low energy.
The renormalization group running will amplify these operators at the
infrared. From a phenomenological perspective these operators have
considerable impact on the phenomenology of these models.  They may
lead to novel signals at colliders \cite{Datta:2012xy, Datta:2012tv},
thus providing a way to probe these models at the LHC.

We introduce a minimal correction to the fermionic and Yukawa sector
of UED Lagrangian given in Eq.~\ref{uedaction}, as given by,

\begin{eqnarray}
\Delta S &=& \int_{-\pi R/2}^{\pi R/2} dx_5 \int d^{4}x~\Bigg\{ r_Q \Big(
\delta(x_5-\pi R/2) + \delta(x_5+\pi R/2) \Big) \Big[ i \overline{Q}_3
  \!\not\!{D} \, Q_3 +i \overline{u}_3 \!\not\!{D} \, u_3 \Big] \Bigg.\nonumber \\ & & - \Bigg. \Big( r_Y
(\delta(x_5-\pi R/2) + \delta(x_5+\pi R/2)) \Big) \times \Big[
  i\lambda^{u}_{5} \overline{Q}_3 u_3 \tilde{H} + {\rm h.c.}  \Big] \Bigg\}.
 \label{blktaction}
\end{eqnarray}

Though originating from radiative corrections, in this letter we will
consider the size $(r_Q,r_Y)$ of the brane operators at the weak scale
to be free and independent parameters.  Inclusion of these terms can
lead to deviations in the bulk profile, couplings and masses of the
bulk fields. Note that 5d gauge invariance
imply universal BLKT parameters. The genesis of this
lies in the observation that for general BLKT parameters the zero mode
bulk profile of the massive gauge bosons are non-flat\cite{Flacke:2008ne}.
To avoid 
such complication we have taken same BLKT parameter for the doublet
and the singlet states in Eq.~\ref{blktaction}.The KK masses obtained by
performing the KK decomposition and solving the equations of motion
with appropriate boundary conditions now shift from their UED
predictions, and are given by the subsequent roots of the following
transcendental equations,
\begin{eqnarray}
  r_Q m^{(n)}= \left\{ \begin{array}{rl}
         -\tan \left(\frac{m^{(n)}\pi R}{2}\right) &\mbox{for $n$ even,}\\
          \cot \left(\frac{m^{(n)}\pi R}{2}\right) &\mbox{for $n$ odd,}
          \end{array} \right.
 \label{massblkt}         
 \end{eqnarray}

In the present context, the Yukawa sector that couples the fermions with 
non-trivial bulk profile with the zero mode of the Higgs that is assumed t
o be essentially flat in the bulk, is of interest and can be written as,

\begin{equation}
 S_{Yukawa}= - \frac{v}{\sqrt{2}} \lambda \int d^4 x \Bigg\{
 \left(\overline{Q}^{(0)}_{L} u^{(0)}_{R} + r'_{Qnn} \overline{Q}^{(n)}_{L} u_{R}^{(n)}
- R'_{Qnn} \overline{u}^{(n)}_{L} Q_{R}^{(n)}+\ldots \right) +~ \mbox{h.c.} \Bigg\},
\label{blktyukawa}
\end{equation}
where, $ r'_{Qnn} (r_Q,r_Y,m^{(n)})$ and $R'_{Qnn}(r_Q,r_Y,m^{(n)})$
are overlap integrals obtained by introducing the bulk profile of the
fermions into Eq.~\ref{blktaction} and integrating the fifth
dimension. The  effects of possible KK level mixing terms
in Eq.~\ref{blktyukawa} will be discussed seperately.
The overlap inegrals can be adopted from the expression given in
\cite{Datta:2012tv} and are given by,
\begin{eqnarray}
 r'_{Qnn\mbox{(e/o)}} &=& \frac{2r_Q+\pi R}{2r_Y+\pi R}
 \left(\frac{2r_Y+\frac{1}{A_{Q^{(n)}}^2}\left[\frac{\pi R}{2}\pm
     \frac{1}{2m_{Q}^{{(n)}}}\sin (m_{Q}^{{(n)}}\pi
     R)\right]}{2r_Q+\frac{1}{A_{Q^{(n)}}^2} \left[\frac{\pi R}{2}\pm
     \frac{1}{2m_{Q}^{{(n)}}}\sin (m_{Q}^{{(n)}}\pi
     R)\right]}\right),\; \nonumber \\ R'_{Qnn\mbox{(e/o)}} &=&
 \frac{2r_Q+\pi R}{2r_Y+\pi R} \left(\frac{2r_Y(B_{Q^{(n)}})^{2}+
   \frac{1}{A_{Q^{(n)}}^2}\left[\frac{\pi R}{2}\mp
     \frac{1}{2m_{Q}^{{(n)}}}\sin (m_{Q}^{{(n)}}\pi
     R)\right]}{\frac{1}{A_{Q^{(n)}}^2} \left[\frac{\pi
       R}{2}\mp\frac{1}{2m_{Q}^{{(n)}}}\sin (m_{Q}^{{(n)}}\pi
     R)\right]}\right),
\end{eqnarray}
where, the subscript (e/o) represents whether $n$ is even or odd.
$A_{Q^{(n)}} = \sin \left(m_{Q}^{{(n)}}\pi R/2\right)$, for $n$ odd
and $\cos \left(m_{Q}^{{(n)}}\pi R/2\right)$, for $n$ even, similarly
$B_{Q^{(n)}} = \cot \left(m_{Q}^{{(n)}}\pi R/2\right)$, for $n$ odd
and $\tan \left(m_{Q}^{{(n)}}\pi R/2\right)$, for $n$ even and
$m_{Q}^{{(n)}}$ is the $n$-th root of Eq.~\ref{massblkt}.  The mass
matrix for the $n$-th KK excitation of the top quark can be written as,
\begin{equation}
  S_{t^{(n)}} = -\int d^4 x \Bigg\{
\begin{bmatrix} \overline{Q}^{(n)}_3, \ \overline{u}^{(n)}_3 \end{bmatrix}_L
\begin{bmatrix}
                m_{Q_3}^{(n)} & r'_{Qnn} \frac{v}{\sqrt{2}} \lambda_{t} \\
                -R'_{Qnn} \frac{v}{\sqrt{2}} \lambda_{t} & m_{u_3}^{(n)}
\end{bmatrix}
\begin{bmatrix} Q^{(n)}_3 \\ u^{(n)}_3 \end{bmatrix}_R
+ ~\mbox{h.c.} \Bigg\}.
\label{massmatrix}
\end{equation}
In the above expressions we have assumed that in the 4d effective
theory obtained by integrating the fifth dimension, the bulk fields
$Q_{L}(x,x_5)$ and $u_{R}(x,x_5)$ now split into a massless zero
mode and an infinite tower of massive 4d states given by
$(Q^{(0)}_{L},Q^{(n)}_{L},Q^{(n)}_{R})$ and
$(u^{(0)}_{R},u^{(n)}_{L},u^{(n)}_{R})$ respectively.

Note that introduction of the BLKT parameters 
lead to KK level mixing at the leading order in the Yukawa
interactions. However the symmetric nature of the BLKT parameters
as introduced in Eq.~\ref{blktaction} confines the mixing within
the odd or even modes, due to a residual unbroken KK parity
in the theory. The mixing angle between the $n$-th KK level and 
$(n+2l)$-th KK level can be estimated as $\theta_{mix} \sim f(r_Y,r_Q,n,l) m_t R/2l,$
where $f(r_Y,r_Q,n,l)$  schematically represents the corresponding
overlap integrals.
Though the mixing angle is supressed by the  new physics scale ($1/R$), 
 significant level mixing between
zero mode and $n=2$ KK states are possible in certain regions of the parameters space
specially for smaller values of $1/R$ \cite{Datta:2012tv}. 
For the collider implications of this mixing in the top sector see \cite{Datta:new}. 
We now turn our attention to make a careful
study of this phenomenon.
The relevant part of the action can be written as,
\begin{equation}
  S_{t^{(0-2)}} = -\int d^4 x \Bigg\{
\begin{bmatrix}\overline{Q}^{(0)}_3, \overline{Q}^{(2)}_3, \ \overline{u}^{(2)}_3 \end{bmatrix}_L
\begin{bmatrix}
               \frac{v}{\sqrt{2}}\lambda_{t} & 0 &  r'_{Q02} \frac{v}{\sqrt{2}}\lambda_{t} \\
               r'_{Q20} \frac{v}{\sqrt{2}}\lambda_{t} & m_{Q_3}^{(2)} & r'_{Q22} \frac{v}{\sqrt{2}} \lambda_{t} \\
                0 & -R'_{Q22} \frac{v}{\sqrt{2}} \lambda_{t} & m_{u_3}^{(2)}
\end{bmatrix}
\begin{bmatrix} u^{(0)}_3 \\ Q^{(2)}_3 \\ u^{(2)}_3 \end{bmatrix}_R
+ ~\mbox{h.c.} \Bigg\}.
\label{mixing}
\end{equation}
where, 
\begin{equation}
r^{\prime}_{Q20} = \frac{2r_Q+\pi R}{2r_Y+\pi R}\left(\frac{1}{\sqrt{2r_Q+\pi R}}\frac{2\left(r_Y -r_Q \right)}{\sqrt{2r_Q+\frac{1}{A_{Q^{(2)}}^2}\left[\frac{\pi R}{2}+\frac{\sin\left(m_{Q}^{(2)}\pi R\right)}{2m_{Q}^{(2)}}\right]}}\right)
\end{equation}

The overlap integral between the zero and second level, $r^{\prime}_{Q20}$ is symmetric in its last two indices. In Eq~\ref{mixing}, the identification of the  lightest eigenvalue
with the SM top with mass in the range, $m_t = 173.07 \pm 0.52 \pm 0.72$ GeV
\cite{Beringer:1900zz} puts
a severe constraint on the allowed BLKT parameters and consequently constrains
the Higgs couplings that we will now discuss.

In order to obtain the Higgs couplings,
we diagonalize the mass matrix in Eq.~\ref{massmatrix}, that
gives us the physical mass of the $n$-th KK
excitations. One can use this to compute the Higgs cross section and
decay width in the BLKT scenario by using Eqs.~\ref{mastereqn} and
\ref{np}. 
We perform the KK summation numerically and terminate the
procedure at $n=20,$ as the contribution decouples with the KK number
and becomes numerically insignificant beyond this.  Note that the
gauge and scalar parts of the Lagrangian are unaffected by the
introduction of the BLKT action given in Eq.~\ref{blktaction}. For the
gauge-Goldstone sector one can adopt the analytic expression in
Eq.~\ref{uedw}. As indicated above  we take care to include the constraint from 
the mixing effect on the parameter space of the theory. We find that once
the constraint on the top mass is imposed the numerical effect of the mixing 
on the Higgs coupling is insignificant. However we include the leading order 
contribution from the mixing effect by considering the 0-2 KK level mixing.
This can be consistently included into the calculation by replacing the 
contribution of the  second KK top to the sum in Eq.~\ref{np} by the following
expression,

\begin{equation}
 {\mathcal{A}}^{(2)}_{NP} = \frac{4}{3} \left[ 
 \sum_{j=1}^{3} \frac{v}{m_j}\frac{\partial m_j}{\partial v} - 1 \right],  
\end{equation}
where, $m_j$-s are the three eigenvalues of the mass matrix in Eq.~\ref{mixing} 
with lowest eigenvalue identified with SM top mass.

\begin{figure*}
 \centering
 \subfigure{\includegraphics[width=.49\textwidth,keepaspectratio,angle=0]
   {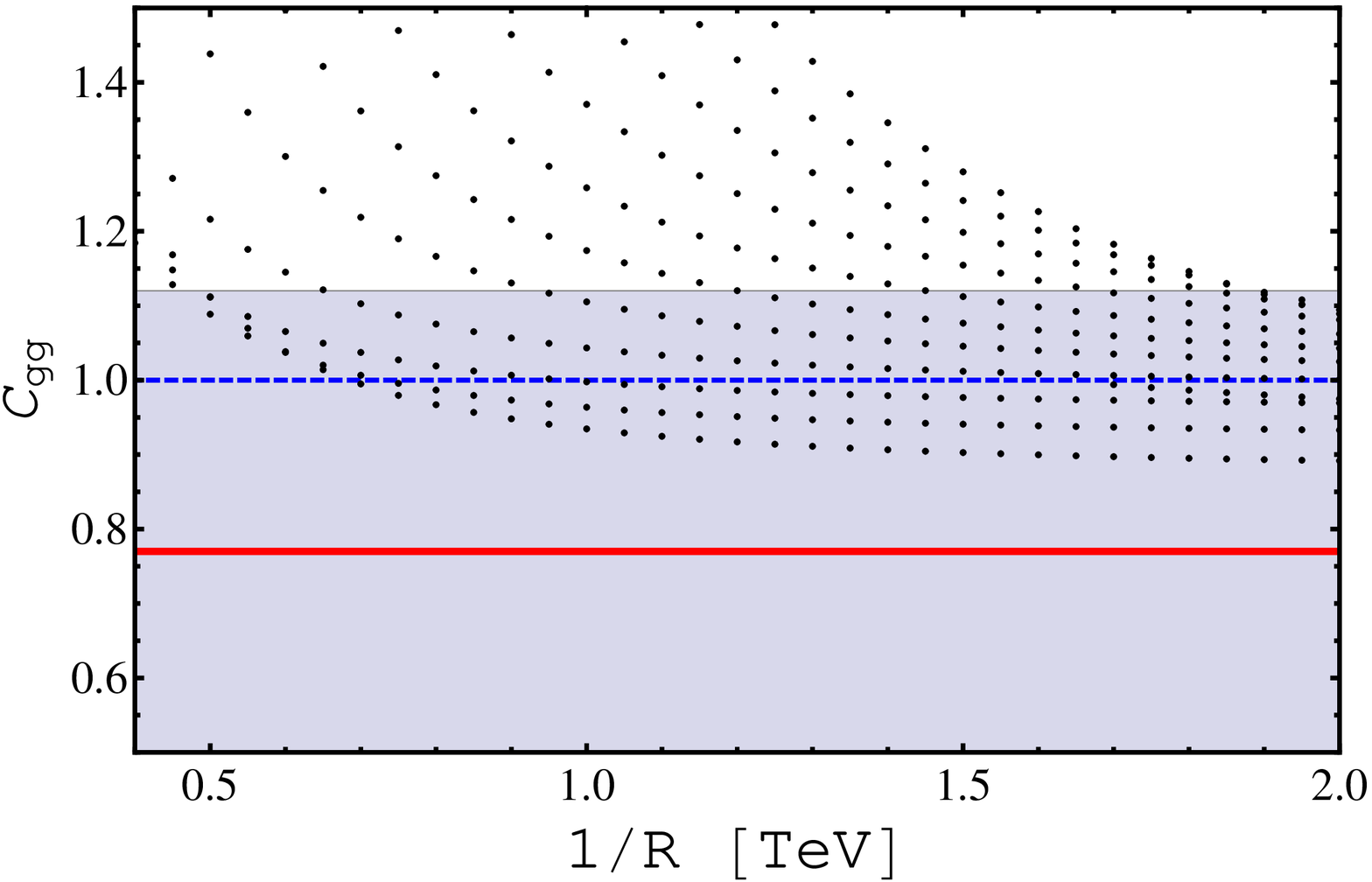}}\subfigure{\includegraphics[width=.49\textwidth,keepaspectratio,angle=0]
   {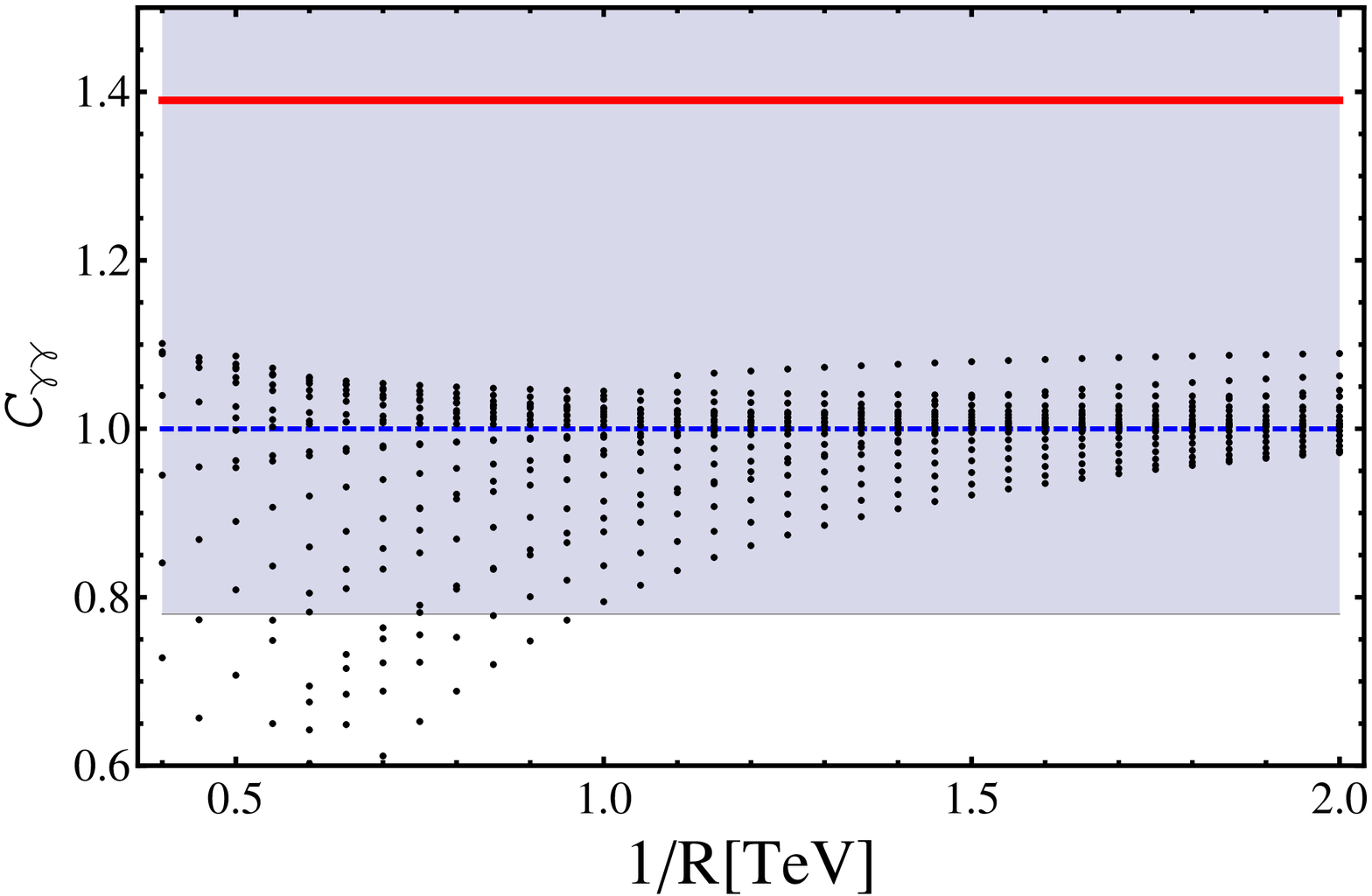}}
             \caption{ The ratios the Higgs couplings in BLKT scenario
               to their SM values are plotted as  functions of the
               inverse radius of compactification of the extra
               dimension (1/R). The gray (shaded) bands represent the
               95\% confidence level best fitted values for these
               ratios, with the red (light) horizontal lines
               representing the central values from LHC data. The blue
               (dashed) lines correspond to the SM points. The black
               (dark) points represent the  BLKT results. We have assumed $m_h = 125$ GeV. The BLKT parameters $r_{Q}$ and $r_{Y}$ are varied within the range $[-\pi R/2, \pi R/2]$.}
            \label{BLKT}
\end{figure*}

After the KK sum is done the ratios $C_{gg}$ and
$C_{\gamma \gamma}$ still remain functions of the BLKT parameters
$(r_Q, r_Y)$ and the inverse radius of compactification $(1/R).$ We
vary the BLKT parameters within the range $[-\pi R/2,\pi R/2]$ and
obtain the corresponding scatter plots for $C_{gg}$ and $C_{\gamma
  \gamma}$ as functions of $1/R$ in Figure~\ref{BLKT}. Expectedly the
points form a band around the minimal UED prediction that corresponds
to $r_Q=r_Y=0.$

Crucially we find that in certain regions of the parameter space, the
contribution from the KK fermions can change sign relative to the zero
mode (SM) contribution.  As can be seen from the plot this reduces
constraint from the coupling, which was at $1.3$ TeV for minimal UED
models, now becoming $0.4$ TeV.  Significantly we find that the UED
couplings are always disfavored over the SM predictions in all regions
of the parameter space by the experimental data. In the extended
scenario with the BLKT parameters, we find that in certain regions of
the parameter space the couplings fit better than the SM.  The limits
on $1/R$, from dark matter relic density measurements and their direct
detection at experiments within the BLKT framework, are rather model
dependent and can lead to considerable relaxation over the minimal UED
bound \cite{Flacke:2013pla, Datta:2013nua}. The corresponding
constraints on this class of models from electroweak precision
measurements can be found in \cite{Flacke:2008ne, delAguila:2003gv,
  Flacke:2013pla}.

\textbf{Conclusion:} In this paper we have studied the impact of Higgs
couplings as measured at the LHC on Universal Extra Dimension
models. We find that the minimal models are particularly constrained
from the Higgs coupling to the gluon. We make a simple extension of
the model by introducing relevant brane localized kinetic terms. This
leads to non-trivial 5d profiles for the bulk fields. The interactions
are modified by the corresponding overlap integrals. In certain
regions of the parameter space this can lead to better fitting of the
experimental data implying a considerable relaxation of the
experimental constraints.

\noindent{ {\bf{Acknowledgments:}} We thank A. Raychaudhuri and Kenji Nishiwaki
  for
  discussions at various stages of this work. UKD would like to thank
  S. Niyogi and A. Shaw for useful discussions. TSR acknowledges
  hospitality at Department of Physics, Calcutta University, during
  the early stages of the work. The research of TSR is supported by
  the Australian Research Council. UKD is supported by funding from
  the Department of Atomic Energy, Government of India for the
  Regional Center for Accelerator-based Particle Physics, Harish-
  Chandra Research Institute (HRI).}

\end{document}